\documentclass[
    ,final            
  ]
  {aipproc}

\layoutstyle{6x9}

\begin{document}

\title{Exclusive production of meson pairs \\
and resonances in proton-proton collisions
\footnote{This work was supported in part by the MNiSW grant No. PRO-2011/01/N/ST2/04116.}}


\author{Piotr Lebiedowicz
}{
  address={Institute of Nuclear Physics PAN, PL-31-342 Cracow, Poland}
}

\author{Antoni Szczurek}{
  altaddress={University of Rzesz\'ow, PL-35-959 Rzesz\'ow, Poland},
  address={Institute of Nuclear Physics PAN, PL-31-342 Cracow, Poland}
}


\begin{abstract}

We report a study of the central exclusive production
of $\pi^{+} \pi^{-}$ and $K^{+}K^{-}$ pairs
in high energy hadron-hadron collisions.
The amplitude is calculated in the Regge approach
including both pomeron and secondary reggeon exchanges
and absorption effects due to proton-proton interaction and $\pi\pi$ ($KK$) rescattering.
We discuss a measurement of exclusive production
of a scalar $\chi_{c0}$ meson via $\chi_{c0} \to \pi^{+} \pi^{-}$, $K^{+} K^{-}$ decay.
We find that the relative contribution of resonance states and
the $\pi\pi$ ($KK$) continuum strongly depend on the cut on pion (kaon) transverse momentum.
We compare the results with the existing experimental data
and present predictions for the RHIC, Tevatron and LHC colliders.
We discuss also the $f_{2}(1270)$ meson production mediated by 
an effective tensor pomeron exchanges.

\end{abstract}

\maketitle


\section{Introduction}
\label{intro}
Central exclusive production (CEP) processes of the type
$pp \to p X p$, where $X$ represents
the centrally produced state separated from the two very forward protons
by large rapidity gaps, 
provide a very promising way to study the properties of resonance states.
We have studied recently the four-body 
$pp \to pp \pi^{+} \pi^{-}$ \cite{SL09,LSK09,LS10}, $pp K^{+}K^{-}$ \cite{LS11_kaons}
reactions which constitute an irreducible background to resonance states
(e.g. $\phi$, $f_{2}(1270)$, $f_{0}(1500)$, $f_{2}'(1525)$, $\chi_{c0}$).
As discussed in \cite{LPS11,LKRS12}, the measurement of $\chi_{c0}$
CEP via two-body decay channels to light mesons 
is of special interest for both studying the dynamics
of heavy quarkonia and for testing the QCD framework of CEP.

CEP processes have been successfully observed at the Tevatron
by selecting events with large rapidity gaps \cite{Albrow}.
At the Tevatron the measurement of exclusive production of $\chi_{c}$
via decay in the $J/\psi + \gamma$ channel cannot provide production
cross sections for different species of $\chi_{c}$ \cite{CDF}.
It may be possible to isolate the $\chi_{cJ}$ CEP contribution 
via hadronic decay channels, especially to 
$\pi^{+} \pi^{-}$ \cite{LPS11} and $K^{+} K^{-}$ \cite{LS11_kaons}.
In particular the branching fraction to these channels 
are relatively larger for scalar meson than for the tensor meson 
and $\sigma(\chi_{c0}) > \sigma(\chi_{c2})$ from theoretical calculation 
means that only $\chi_{c0}$ will contribute to the signal \cite{PST_chic,LPS11}.

A new area of experimental studies of CEP with
tagged forward protons has just started.
It is expected that large CEP data sample
will be available in the near future from measurements
performed by the STAR Collaboration at RHIC \cite{Adamczyk}.
In Ref. \cite{SLTCS11} a possibility of measuring exclusive $\pi^+ \pi^-$ production
at the LHC with tagged forward protons (ALFA detectors) 
during special low-luminosity runs has been studied.
The $pp \to nn \pi^+ \pi^+$ \cite{LS11} and $pp \to pp \omega$ \cite{CLSS11} processes 
are also very interesting for possible future experiments at high energies.

\section{Sketch of formalism}
\label{formalism}

\begin{figure}[!h]
a)  \includegraphics[height=.16\textheight]{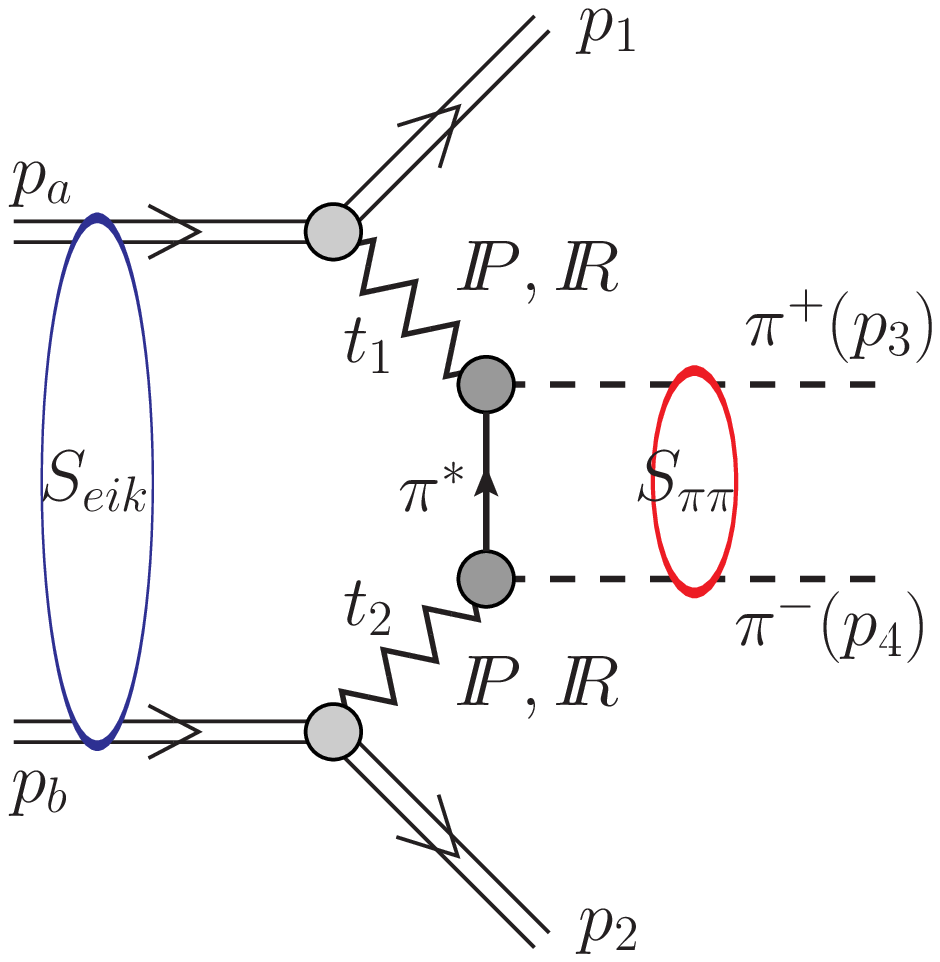}
b)  \includegraphics[height=.14\textheight]{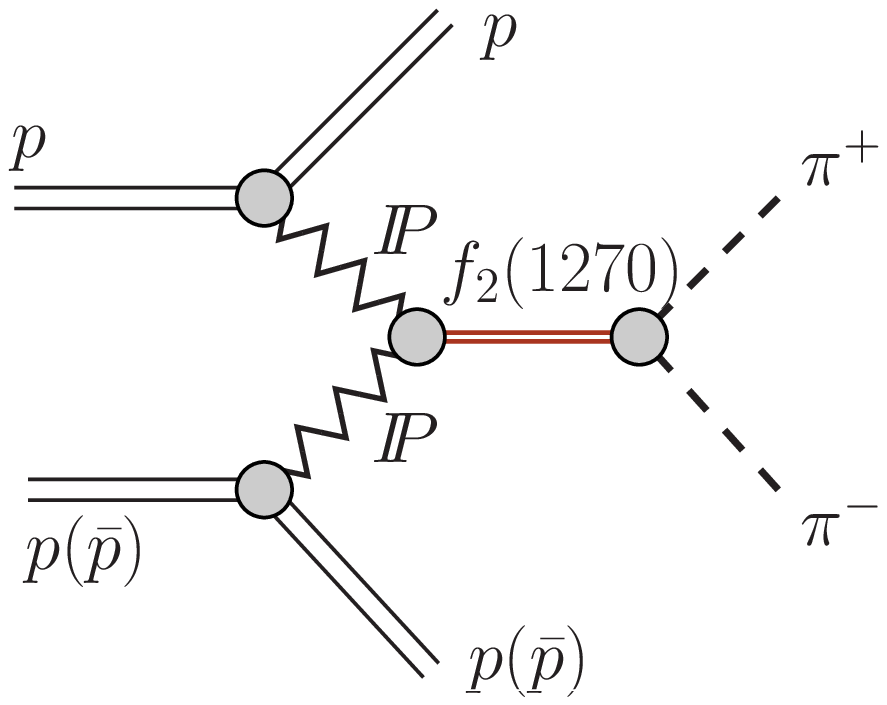}
c)  \includegraphics[height=.15\textheight]{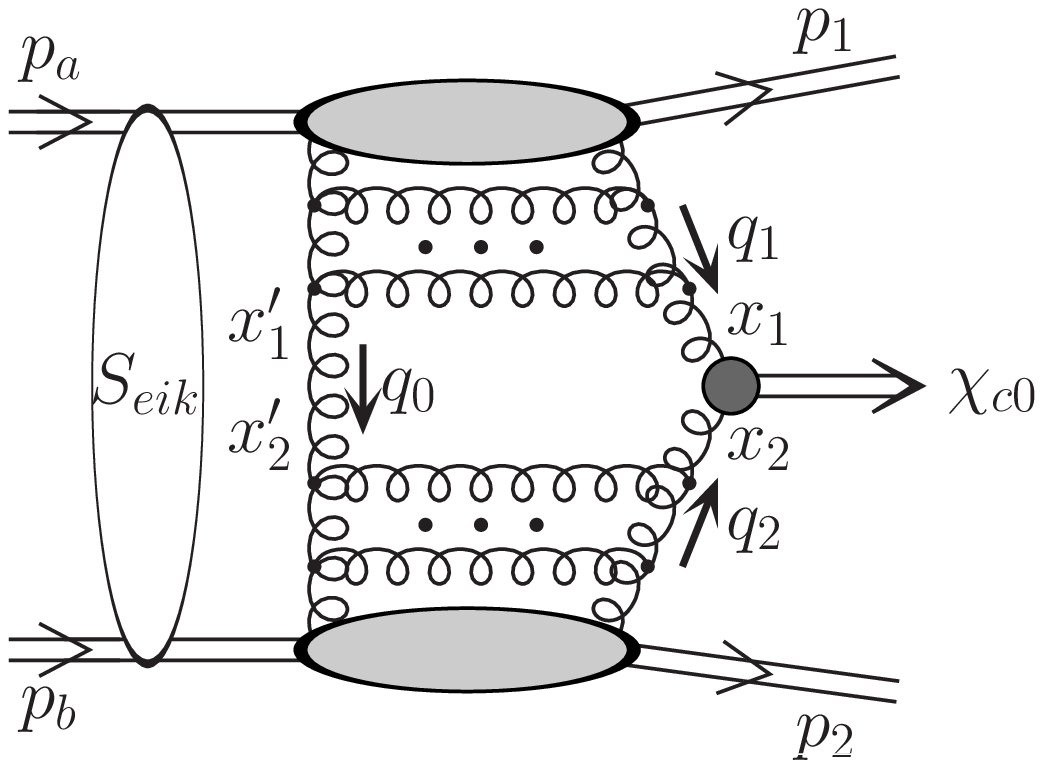}
\caption{
Representative diagrams for the non-perturbative
exclusive production of pion (kaon) pairs (panel a) and $f_{2}(1270)$ meson (panel b).
The absorptive corrections due to proton-proton interactions
and $\pi\pi$-rescattering are indicated
and the perturbative mechanism of $\chi_{c0}$ meson production (panel c).
}
\label{fig:1}
\end{figure}

The dominant mechanism of the exclusive production of
light meson pairs at high energies is sketched in Fig.\ref{fig:1}a.
The formalism used to calculate
of non-resonant background amplitude is explained in detail in Refs \cite{LS10,LS11_kaons}.
The Regge parametrization of the scattering amplitude includes 
both pomeron and subleading reggeon exchanges.
Our model with the parameters taken from the Donnachie-Landshoff analysis
of the total $\pi N$ or $KN$ cross sections
sufficiently well describes the elastic data for $\sqrt{s} > 3$ GeV.
The form factors correcting for the off-shellness of the intermediate pions/kaons
are parametrized as
$F_{\pi/K}(\hat{t}/\hat{u})=
\exp\left(\frac{\hat{t}/\hat{u}-m_{\pi/K}^{2}}{\Lambda^{2}_{off}}\right)$,
where the parameter $\Lambda_{off}^{2} = 2$ GeV$^{2}$ is obtained from a fit 
to the CERN-ISR data \cite{ABCDHW89,ABCDHW90}.

In Fig.\ref{fig:1}b the exclusive dipion production through 
the $s$-channel $f_{2}$-meson exchange and double tensor pomeron exchange 
is presented \cite{LNS_preparation}. 
The theoretical arguments for an effective tensorial answer for
the nonperturbative pomeron are sketched in \cite{talkN}
and will be discussed in detail \cite{EMN_preparation}.

The QCD amplitude for exclusive central diffractive $\chi_{c0}$ meson production,
sketched in Fig.\ref{fig:1}c,
was calculated within the $k_{t}$-factorization approach
including virtualities of active gluons \cite{PST_chic}
and the corresponding cross section is calculated with the help of
unintegrated gluon distribution functions (UGDFs).
In Ref.\cite{LPS11} we have performed detailed studies of several differential distributions of $\chi_{c0}$ meson production.

\section{Results}
\label{results}

\begin{figure}[!h]
  \includegraphics[height=.22\textheight]{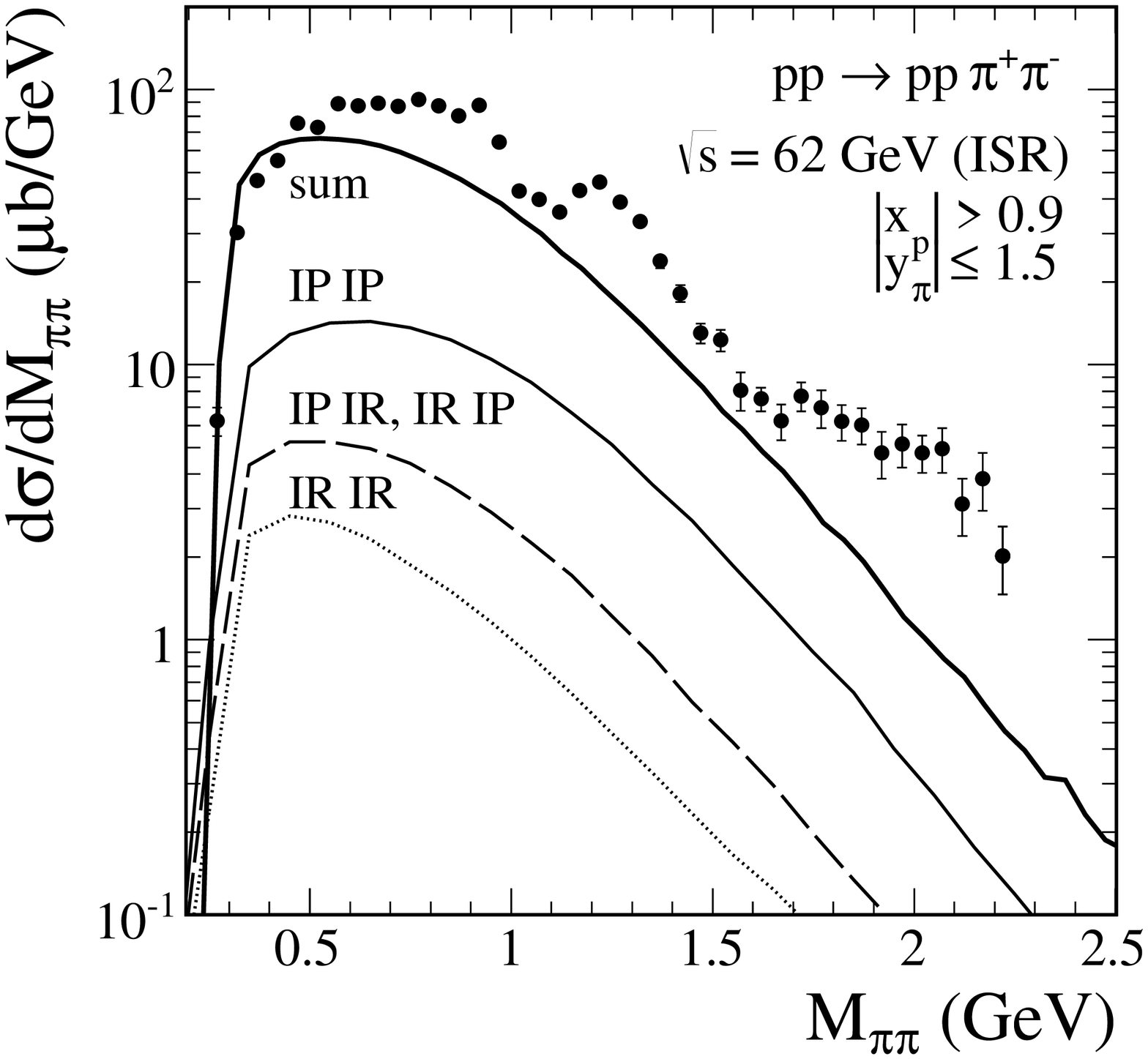} 
  \includegraphics[height=.22\textheight]{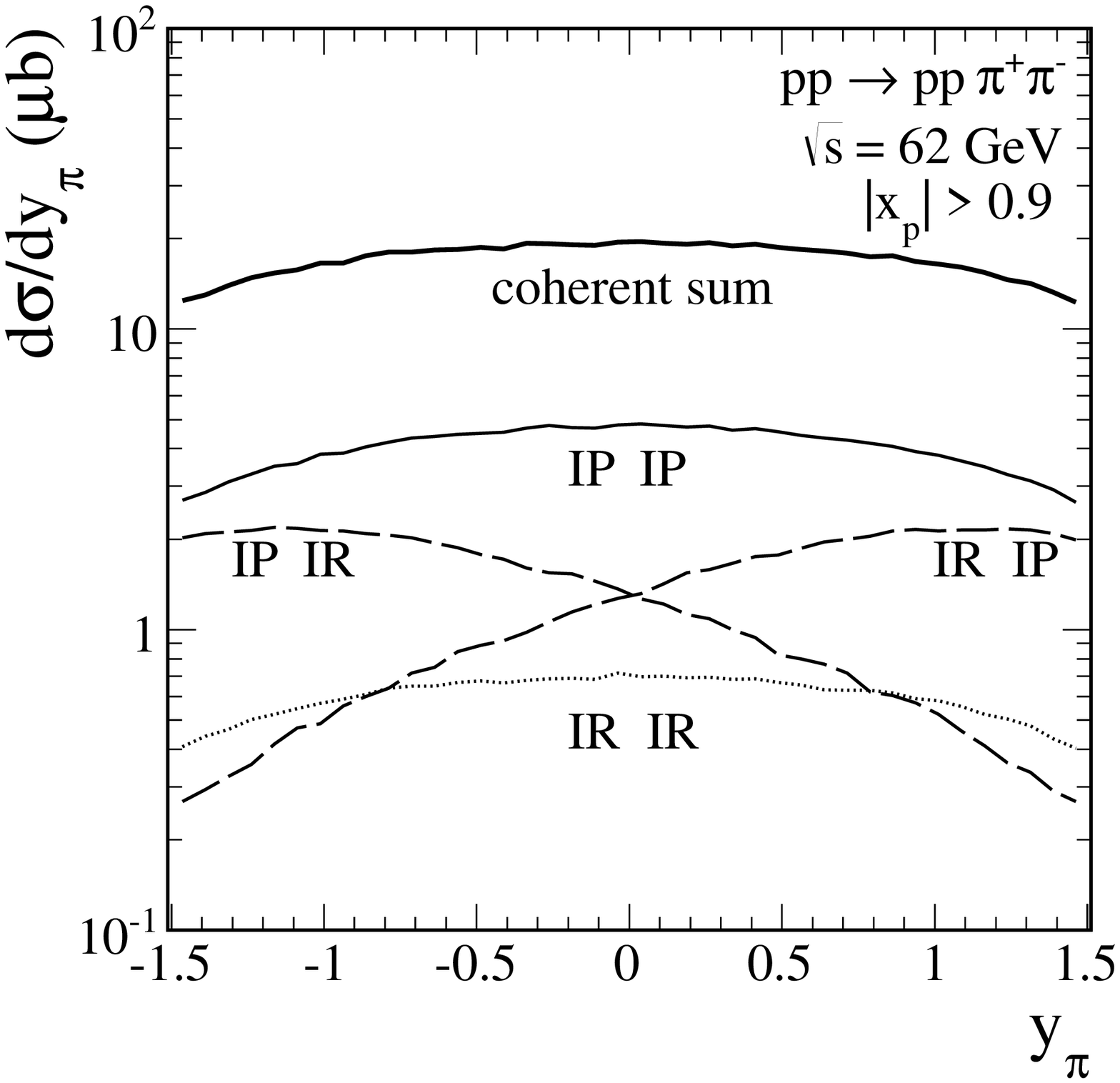} 
  \includegraphics[height=.22\textheight]{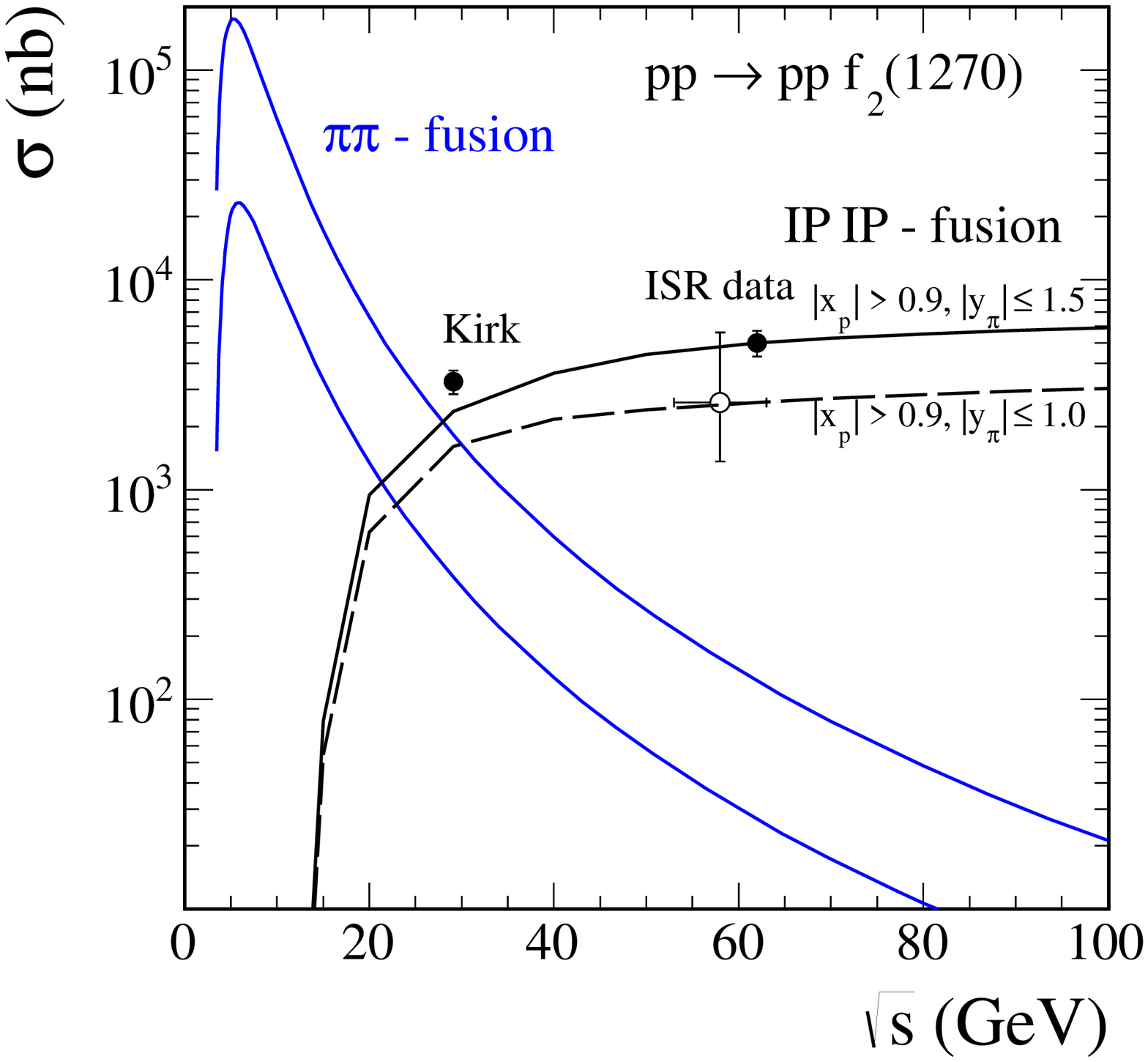}
\caption{Left panel:
Differential cross section in $\pi^{+}\pi^{-}$ invariant mass and pion rapidity
at $\sqrt{s} = 62$ GeV with experimental cuts relevant for the ISR data \cite{ABCDHW90}.
The pomeron-pomeron component dominates at midrapidities of pions 
and pomeron-reggeon (reggeon-pomeron)
peaks at backward (forward) pion rapidities, respectively.
Cross section for the $pp \to pp f_{2}(1270)$ reaction as a function
of $pp$ center-of-mass energy.
We see data points from the WA102 \cite{kirk00} and the ISR \cite{ABCDHW90, Waldi} experiments.
We show in addition the cross section for $\pi \pi$-fusion mechanism \cite{SL09}
for two different values of the form factor parameters.
}
\label{fig:2}       
\end{figure}

In Fig.~\ref{fig:2} we compare our results with
CERN ISR experimental data \cite{ABCDHW90} at $\sqrt{s} = 62$ GeV.
One can see two-pion invariant mass spectrum
with strong resonance structures attributed to $f_{0}$ and $f_{2}$ states
and distribution in pion rapidity
when all (solid line) and only some components in the amplitude are included.
In the right panel the cross sections for the $f_{2}(1270)$ meson production
was calculated according to the diagram in Fig.\ref{fig:1}b
with an effective tensor pomeron exchanges \cite{LNS_preparation}.
In principle, the resonance and continuum contributions should be added
coherently together leading to the distortion of the $f_{2}$
line shape as observed e.g. for the $\gamma \gamma \to f_{2}(1270) \to \pi^{+} \pi^{-}$ 
reaction \cite{SS2003}.
This requires a consistent model of the resonances and the backgrounds.
%

\begin{figure}[!h]
  \includegraphics[height=.21\textheight]{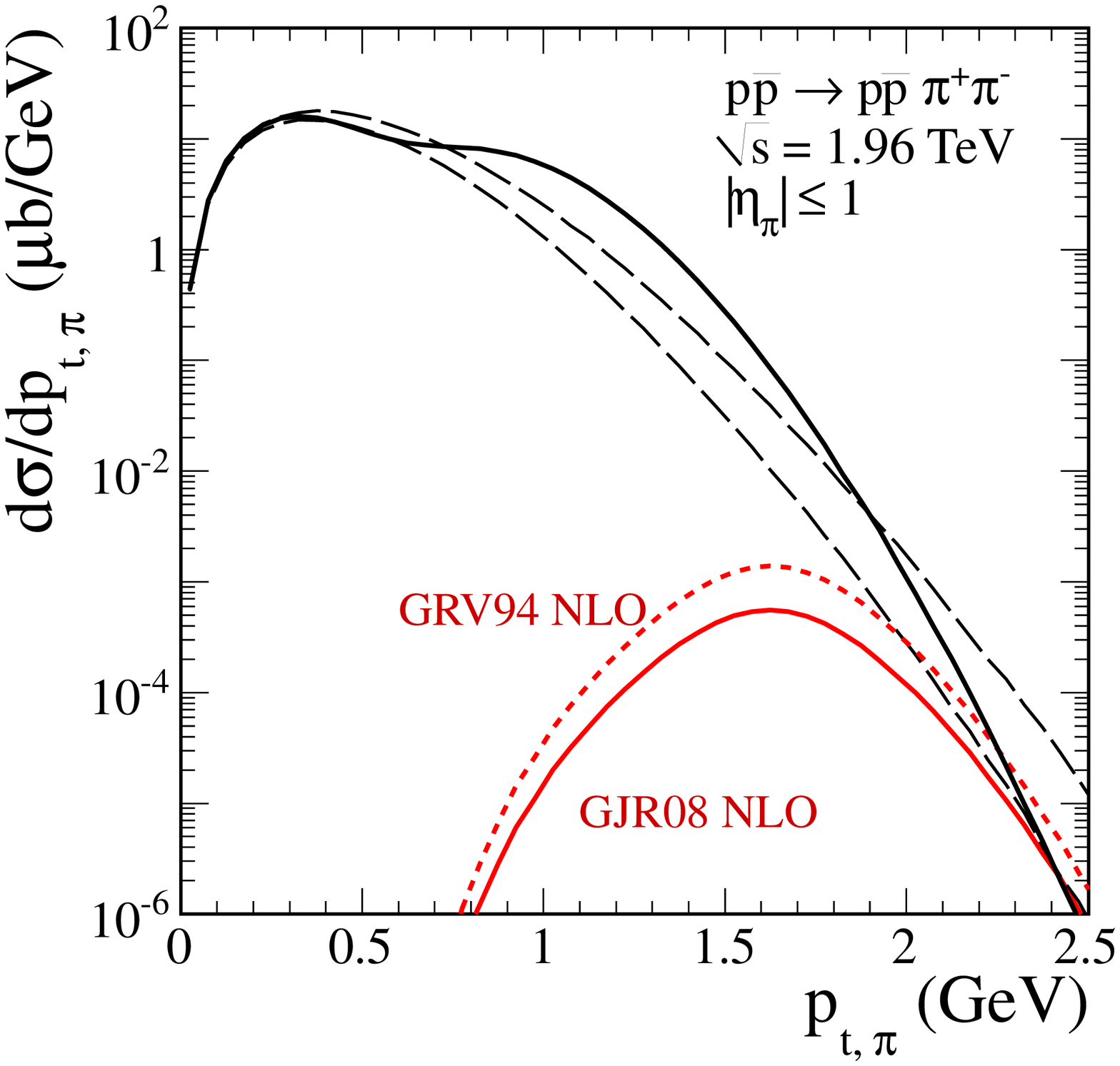}
  \includegraphics[height=.21\textheight]{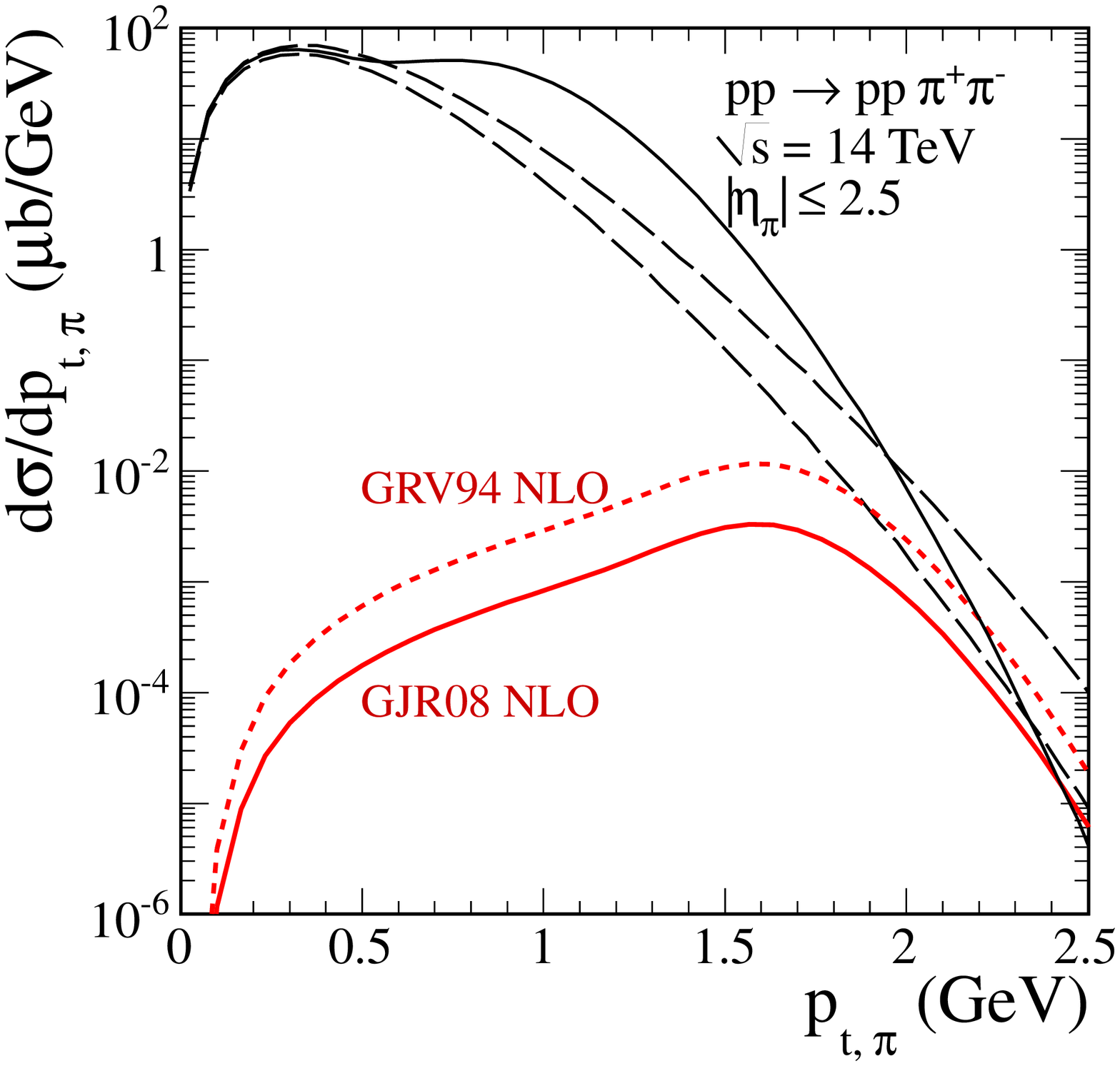}
  \includegraphics[height=.21\textheight]{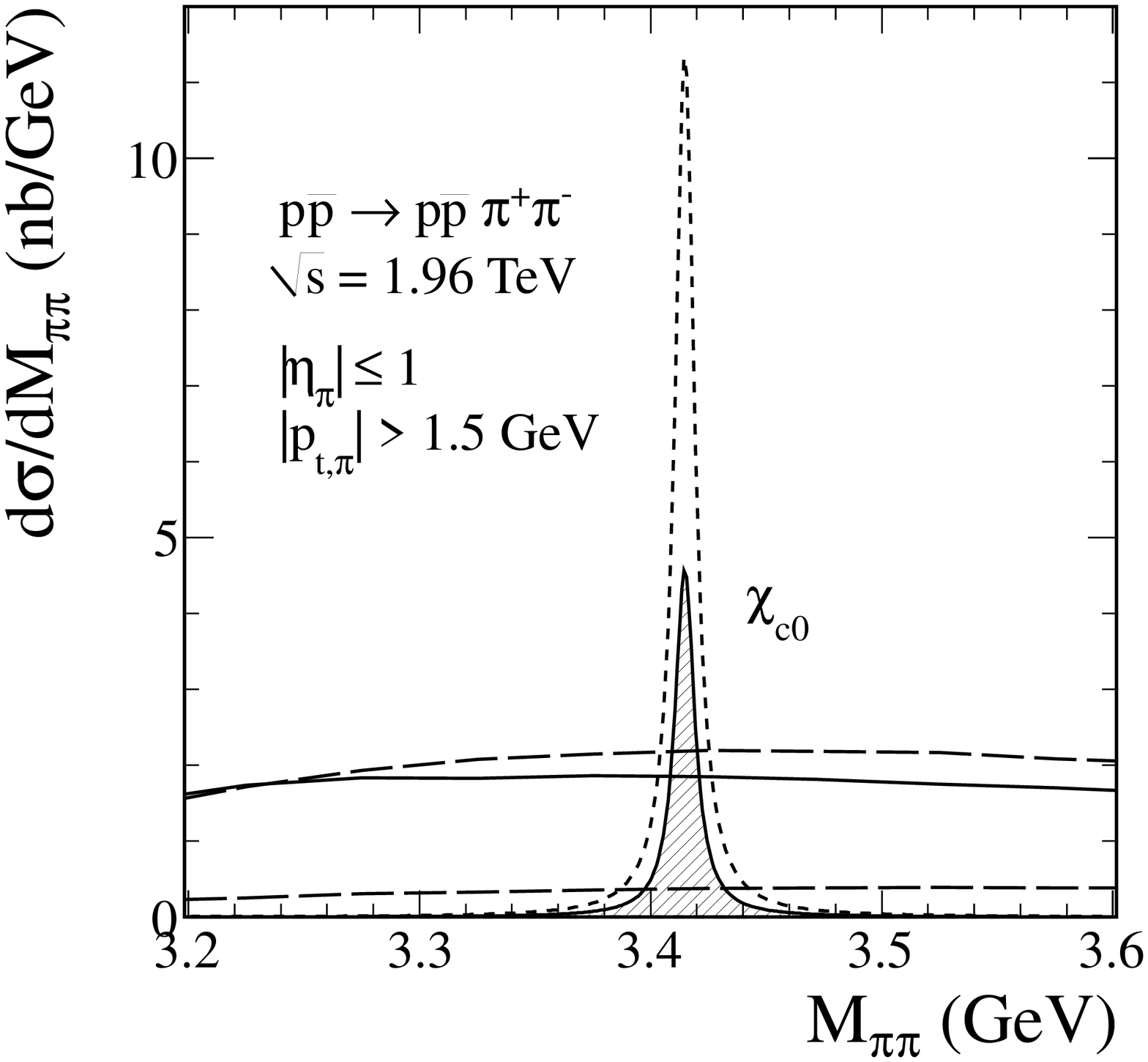}
  \includegraphics[height=.21\textheight]{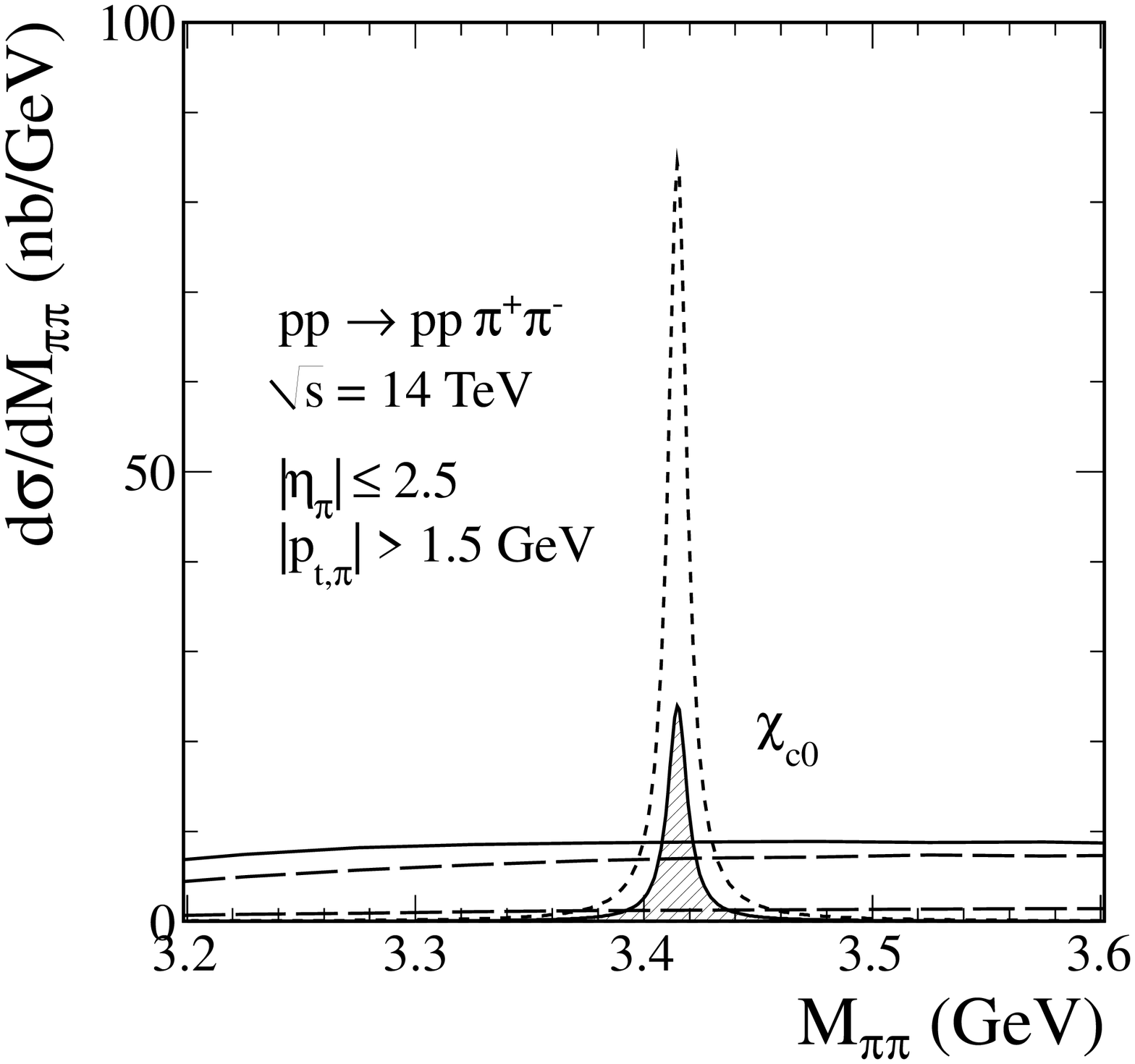}
\caption{
Differential cross section $d\sigma/dp_{t,\pi}$ and $d\sigma/dM_{\pi\pi}$ at $\sqrt{s} =
1.96, 14$ TeV with cuts on the pion pseudorapidities.
Results for the $\pi\pi$ continuum with the meson propagator
and with the cut-off parameter $\Lambda_{off}^{2}$ = 1.6, 2 GeV$^{2}$ 
(lower and upper dashed lines, respectively) as well as
with the generalized pion propagator and $\pi\pi$-rescattering effect (solid line) 
are presented.
In the calculation of the $\chi_{c0}$ distributions we have used 
GRV94 NLO \cite{GRV} (dotted lines) and GJR08 NLO \cite{GJR} (filled areas) collinear gluon distributions.
An additional cuts on both pion transverse momenta
$|p_{t,\pi}| > 1.5$ GeV improve significantly the S/B ratio (two right panels).
The cuts play then a role of the $\pi \pi$ (or $KK$) resonance filter.
}
\label{fig:3}       
\end{figure}

In Fig.~\ref{fig:3} the $p_{t,\pi}$ and $M_{\pi\pi}$ distributions
both for the signal ($\chi_{c0}$) and background are presented.
The absorption effects were included in the calculations.
The fact that pions from the $\chi_{c0}$ decay are placed at larger $p_{t,\pi}$
can be used to improve the signal-to-background ratio \cite{LPS11}.
Measurements of other decay channels, e.g. $K^{+}K^{-}$, are possible as well \cite{LS11_kaons}.

\section{Conclusions}
We have calculated several differential observables for the
$pp \to pp \pi^{+} \pi^{-}$ \cite{SL09,LSK09,LS10} and
$pp K^{+}K^{-}$ \cite{LS11_kaons} reactions.
The full amplitude of central diffractive process
was calculated in a simple model with parameters adjusted to low energy data.
At high energies the pions or kaons from the presented CEP mechanism
are emitted preferentially in the same hemispheres, 
i.e. $y_{\pi^{+}}, y_{\pi^{-}} >$ 0 or $y_{\pi^{+}}, y_{\pi^{-}} <$ 0.
We have predicted large cross sections for RHIC, Tevatron and LHC
which allows to hope that presented by us distributions will be measured in near future.
We have calculated also contributions of several diagrams
where pions/kaons are emitted from the proton lines.
These mechanisms contribute at forward and backward regions
and do not disturb the observation of the central DPE component 
which dominates at midrapidities.

We have analyzed a possibility to measure the
exclusive production of $\chi_{c0}$ meson in the proton-(anti)proton
collisions at the RHIC, Tevatron and LHC
via $\chi_{c0} \to \pi \pi$, $KK$ decay channels.
For a more detailed discussion of this issue see \cite{LPS11,LS11_kaons,LKRS12}.

Future experimental data on exclusive meson production at higher
energies may provide a better information on the spin structure of 
the pomeron and its coupling to the nucleon and mesons. 
The relevant measurements at high energies are possible
and could provide useful information 
e.g. about $f_{0}(980)$, glueball candidate $f_{0}(1500)$ \cite{SL09}, 
$f_{2}(1270)$ and $\chi_{c0}$ meson CEP production.




\bibliographystyle{aipproc}   

\end{document}